# Extraordinary sensitivity of the electronic structure and properties of single-walled carbon nanotubes to molecular charge-transfer

Rakesh Voggu [a], Chandra Sekhar Rout [a], Aaron D. Franklin [b], Timothy S. Fisher [b] and C. N. R. Rao[a*]

[a] Chemistry and Physics of Materials Unit, Jawaharlal Nehru Centre for Advanced Scientific Research, Jakkur P.O., Bangalore -560 064, India.
E-mail: cnrrao@jncasr.ac.in Fax: (+91)80-22082766

[b] Brick Nanotechnology Center, Purdue University, West Lafayette, Indiana 47907, USA.

**Summary:**

Interaction of single-walled carbon nanotubes with electron donor and acceptor molecules causes significant changes in the electronic and Raman spectra, the relative proportion of the metallic species increasing on electron donation through molecular charge transfer, as also verified by electrical resistivity measurements.

---

* For correspondence: cnrrao@jncasr.ac.in

Single-walled carbon nanotubes (SWNTs) exhibit diverse electronic structure and properties arising from the quantization of electron wave vector of the 1D system.[1-3] Thus, SWNTs exhibit significant changes in the electronic structure and chemical reactivity depending on the geometry, doping, chemical environment and solvent. Electron or hole doping influences the electronic structure of SWNTs and thereby their Raman spectra [4] and electrochemical doping also causes similar effects.[5] Interaction of SWNTs with gold and platinum nanoparticles can give rise to a semiconductor to metal transition.[6] More interestingly, aromatic solvents containing electron donating and electron withdrawing groups have been shown to modify the electronic structure of the nanotubes giving rise to changes in the electrical properties.[7] Clearly the electron transfer caused by the interaction of SWNTs with other molecules depends on the density of states near the Fermi level.[8] We have investigated the interactions of electron-withdrawing and electron-donating molecules with SWNTs to determine the sensitivity of their electronic structure and properties to molecular charge-transfer. The electron donor and acceptor molecules considered here include aniline, tetrathiafulvalene (TTF), nitrobenzene, tetracyanoquinodimethane (TCNQ) and tetracyanoethylene (TCNE). We have made use of both electronic and Raman spectroscopy, besides the measurement of electrical conductivity to study the interaction of SWNTs with these electron donor and acceptor molecules.

SWNTs prepared by arc discharge using the $Y_2O_3$+Ni catalyst were purified by acid and hydrogen treatment following a previously reported procedure [9]. In order to study the interaction of nitrobenzene and aniline, the SWNTs were soaked in the corresponding liquid placed on a glass slide for 1 hour. The solids obtained after drying were examined by electronic absorption and Raman spectroscopies. Electronic spectra were recorded with thin films of SWNTs on a quartz plate. In order to study the interaction with TCNQ, TCNE and TTF, these compounds were added to a suspension of SWNTs in toluene. The suspensions were drop coated on glass or quartz plates and dried in vacuum. In order to compare the effects of these different donor and acceptor molecules, we employed interaction with benzene as a reference. Raman spectra were recorded with a LabRAM HR high-resolution Raman spectrometer (Horiba-Jobin Yvon) using a He−Ne laser ($\lambda$ = 632.8 nm) on a clean glass slide. Electronic absorption spectra were recorded with a UV/VIS/NIR Perkin–Elmer spectrometer.

In Fig. 1, we show the $S_{22}$ bands of SWNTs in the electronic absorption spectra. On interaction with nitrobenzene and aniline, the $S_{22}$ band shifts in the opposite directions relative to the benzene reference. Thus, the positions of the $S_{22}$ band maxima are 1010 and 1090 nm respectively on interaction with nitrobenzene and aniline, while the band peak occurs at 1054 nm on interaction with benzene. Similarly, on interaction with TTF which is a good electron donor, the $S_{22}$ band shifts to 1095 nm while with TCNE and TCNQ band shift to 990 nm and 1005 nm respectively.

Considering that the G-band of SWNTs in the Raman spectrum is highly sensitive to the electronic effects, we have examined the changes in the G-band position on interaction with the electron withdrawing and donating molecules. In Fig. 2, we show the G bands recorded with different chemical surroundings. The G band maximum is shifted to 1582 $cm^{-1}$ and 1590 $cm^{-1}$ respectively in the case of aniline and nitrobenzene, compared to 1585 $cm^{-1}$ with benzene. These shifts are similar to those reported by Shin et a..[7] Carbon nanotubes doped with boron and nitrogen produce shifts in the opposite directions, just as nitrobenzene and aniline. While boron doping (p-type) increases the G band frequency, nitrogen doping (n-type) decreases the G band frequency. We, therefore, surmise that aniline causes n-type doping while nitrobenzene corresponds to p-type doping. Further, the G-band is shifted to 1574 cm$^{-1}$ on interaction with TTF, while with TCNQ and TCNE the G-band maxima occur at 1593 and 1599 cm$^{-1}$ respectively. These shifts are also consistent with the electron- donating or withdrawing character of these chemicals.

The Raman G band of SWNTs exhibits a significant feature around 1540 cm$^{-1}$ due to the metallic species.[5, 7] The Raman G-band can accordingly be deconvoluted to obtain approximate estimates of the metallic species relative to the semiconducting species (Fig. 3). We find that the metallic feature in the G-band appears prominently on interaction of the SWNTs with aniline and TTF, but nearly disappears on interaction with nitrobenzene, TCNQ and TCNE. It appears that donor molecules increase the apparent proportion of the metallic species by making the semiconducting species more conductive through electron donation. Electron withdrawing molecules, on the other hand, decrease the apparent proportion of the metallic species by withdrawing carriers from the metallic species. This conclusion is consistent with the observation of Shin et. al..[7] Thus, the percentage of metallic like species varies as TTF > aniline > benzene > nitrobenzene > TCNQ > TCNE. We estimate the percentage of the metallic species to be less than 10% on interaction with electron-withdrawing molecules, and to reach 35% or more on interaction with electron-donating molecules.

We find that molecular charge-transfer also affects the Raman signature of radial breathing modes (RBM) of SWNTs. Although it is difficult to estimate the proportions of the metallic and semiconducting species by making use of the RBM bands, the occurrence of marked changes in the band shapes due to molecular charge transfer is noteworthy. In Fig. 4, we show the RBM bands of SWNTs recorded after interaction with TTF and TCNE to illustrate this feature.

In order to carry out electrical resistivity measurements vertical SWNT devices were fabricated in a porous anodic alumina (PAA) template[10-13], the process involves the deposition of a thin Ti layer on the surface of the wafer to serve as an electrical back-contact. Three subsequent metal layers, 100 nm Al, 1 nm Fe, and 500 nm Al, are then deposited onto the wafer. The Al layers are used to create a nanoporous template, while the thin middle layer (Fe) ultimately serves as the catalyst. The metal films and substrate are then submerged in an acid and anodized using a standard two-step technique. The anodization process forms self-assembled nanopores in the Al layer as it oxidizes and turns to PAA. After anodization, a microwave plasma chemical vapor deposition (MPCVD) system is used to create hydrogen plasma at a microwave power of 300 W and substrate temperature of 900°C. The substrate is exposed to this plasma for 10 minutes to reductively open the bottom of the oxide layer. Methane gas is then introduced into the MPCVD chamber to provide a carbon source for growth of the nanotubes from the Fe catalyst on the pore wall. During this process, SWNTs grow with a typical density of one per pore. Subsequently, Pd is electrodeposited into the pore bottoms. Continued Pd electrodeposition past the establishment of bottom contacts to SWNTs results in the formation of Pd nanoclusters that concentrically surround SWNTs on the top PAA surface [14]. Many of the clusters exhibited cubic or a combination of cubic and pyramidal geometries.

A thin layer consisting of Ti (~20 nm) and Au (~50nm) was deposited by physical vapor deposition on top of the PAA membrane to achieve the top contact for the electrical resistivity measurements. Bottom contacts were taken from the Ti-coated Si substrates. The I-V characteristics of this SWNT obtained by conducting AFM measurements exhibit a nearly linear behavior, establishing that the electrical contacts were well established. The I-V characteristics of the SWNT bundle prepared as above and soaked in the liquids of electron-withdrawing and electron-donating molecules were measured using a Keithley 236 multimeter.

Fig. 5 presents the I-V characteristics of the SWNTs in air and in the presence of aniline, anisole, chlorobenzene and nitrobenzene, with the inset showing the calculated resistance at a bias voltage of 0.05 V. The resistance of the SWNTs is 492 $\Omega$ in air at a bias voltage of 1 V. At a forward bias, the resistance decreases markedly in the presence of electron withdrawing molecules such as nitrobenzene and chlorobenzene, whereas it increases in the presence of electron donating molecules such as aniline and anisole. Thus, the resistance of the SWNTs was 2.5 k$\Omega$ and 725 $\Omega$, in the presence of aniline and anisole respectively at 1 V, the corresponding values in the presence of the chlorobenzene and nitrobenzene being 297 $\Omega$ and 217 $\Omega$ respectively. At a bias voltage of 0.05 V, the resistance of the SWNTs is 1.0 k$\Omega$ in air. In the presence of aniline, anisole, chlorobenzene and nitrobenzene, the resistance values are 6.6, 1.7, 0.5 and 0.3 k$\Omega$ respectively. SWNTs under ambient conditions usually show p-type behavior.[15] It is possible that in the presence of electron-donating molecules, the number of hole carriers is reduced, causing an increase in the resistance. In the presence of electron-withdrawing molecules more holes are generated in semiconducting SWNTs. Interestingly, that the resistance varies proportionally with the Hammett substituent constants, the order being $NH_2$ > $OCH_3$ > $Cl$ > $NO_2$. This trend reflects the changes in the nature and the concentration of carriers as well as their mobilities brought about by interaction with electron donor and acceptor molecules. The I-V curves become more nonlinear as one goes from aniline to nitrobenzene. The slope of the I-V curve also increases going from nitrobenzene to aniline, due to the presence of a higher proportion of metallic like nanotubes in the presence of aniline.

In conclusion, the present study establishes the high sensitivity of SWNTs to molecular charge-transfer. This feature of SWNTs may be useful in many of the applications such as sensors and nanoelectronics.

## Figure Captions:

Fig. 1: Electronic absorption spectra of SWNTs on interaction with electron donor and acceptor molecules: (a) (1) benzene, (2) aniline and (3) Nitrobenzene; (b) (1) benzene, (2) TTF and (3) TCNE

Fig. 2: G-bands in the of Raman spectra of SWNTs on interaction with electron donor and acceptor molecules.

Fig 3: Deconvolution of the G bands of SWNTs interacting with (a) TTF and (b) TCNE. M and S stands for metallic and semiconducting species respectively.

Fig. 4: RBM bands in the Raman spectra of SWNTs on interaction with (a) TTF and (b) TCNE

Fig. 5: I-V characteristics of the SWNTs in air and in the presence of different aromatic molecules attached with electron-withdrawing and electron-donating groups.

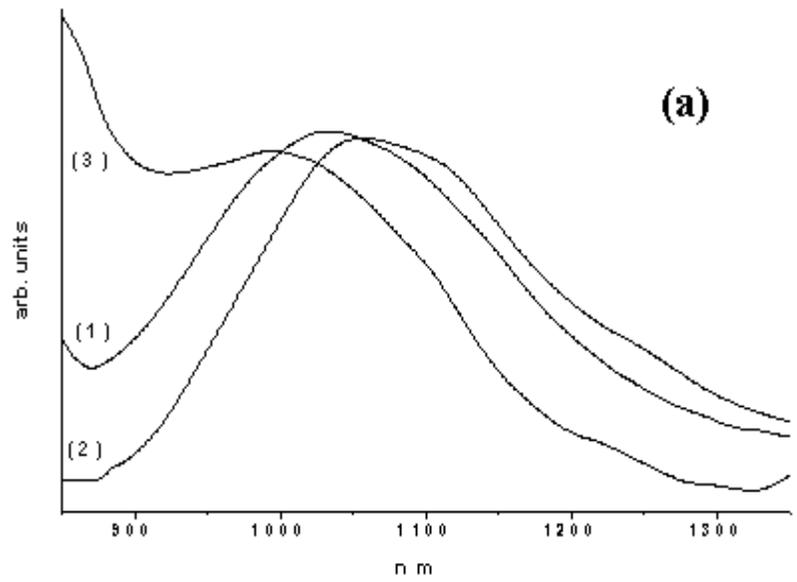
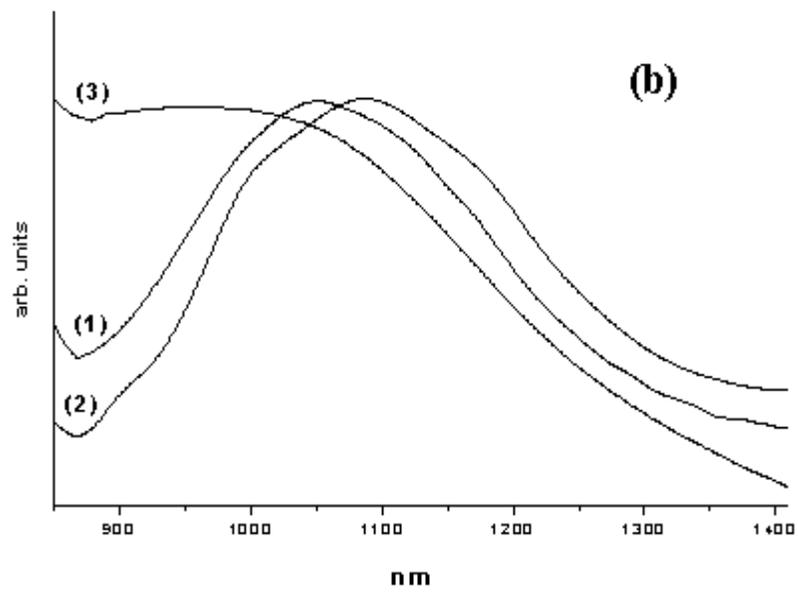

**Fig. 1**

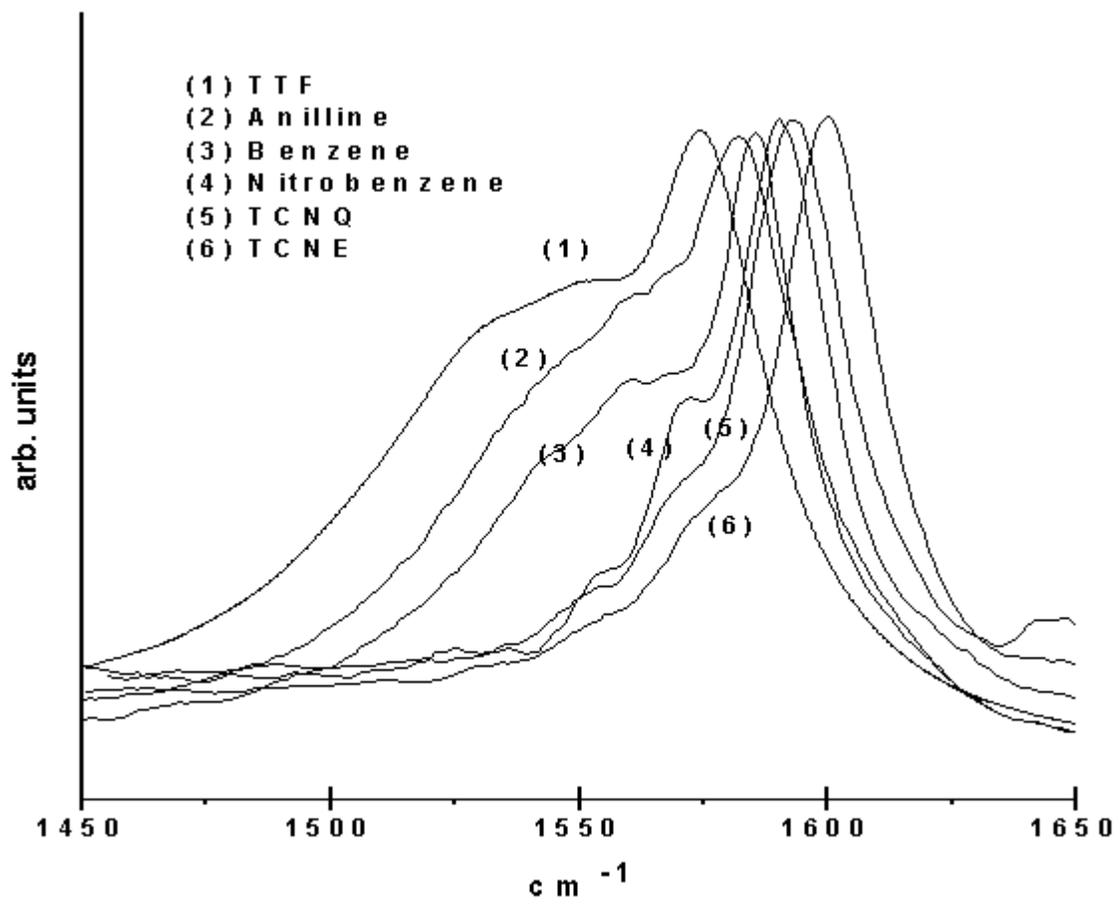

**Fig. 2**

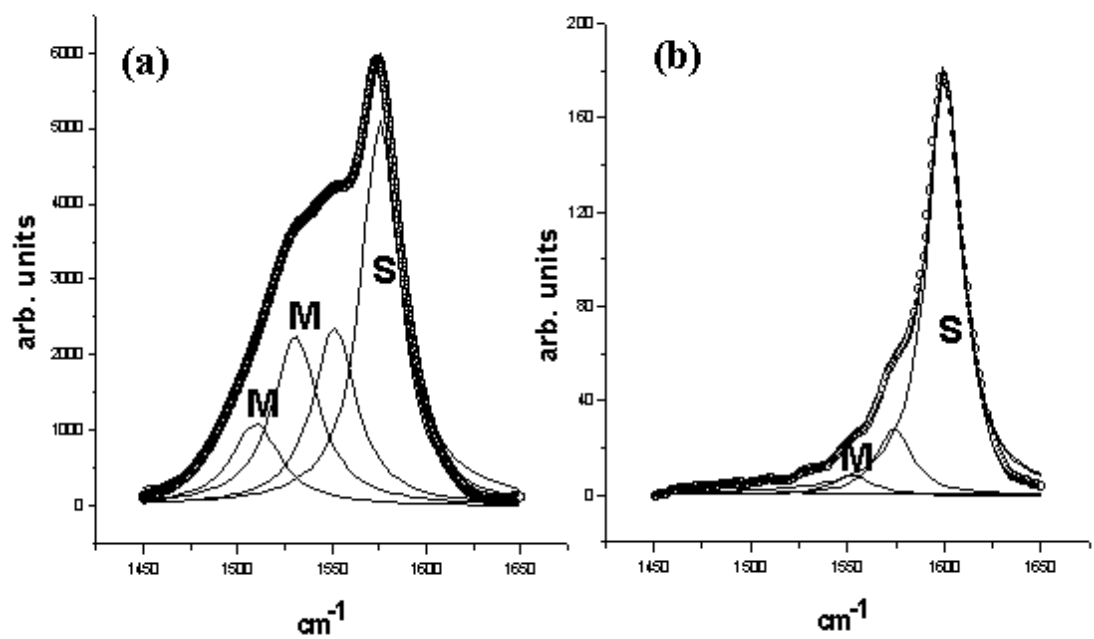

**Fig. 3**

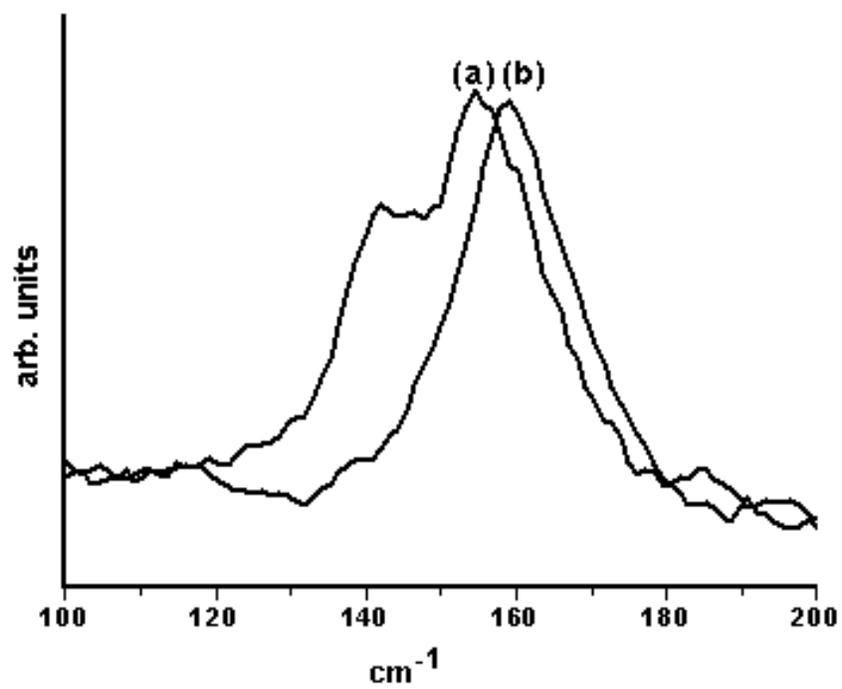

**Fig. 4**

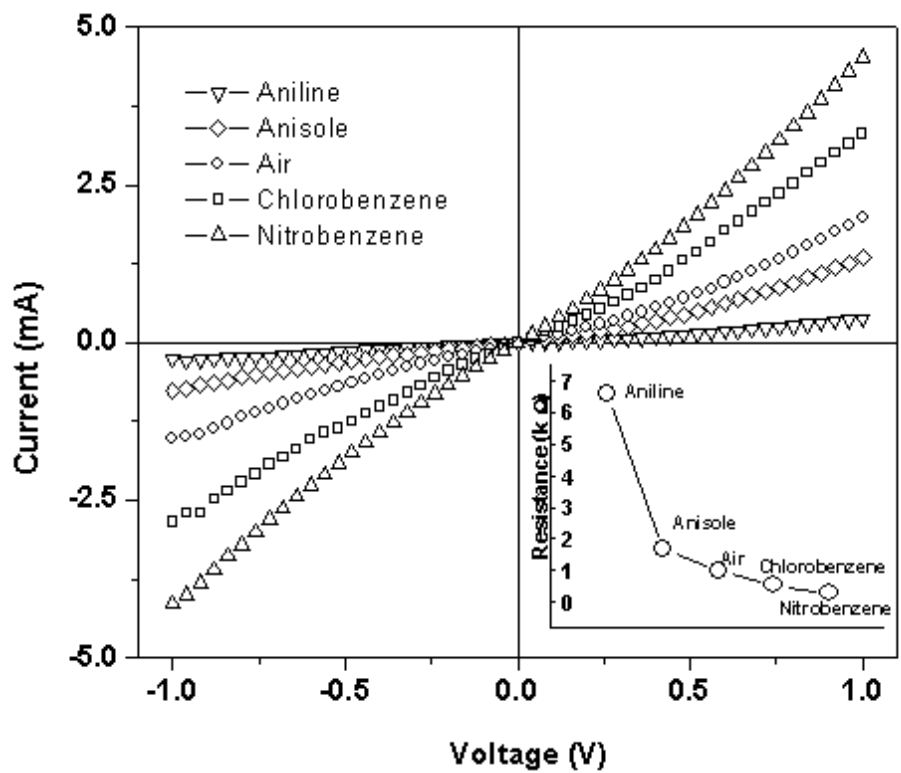

**Fig. 5**